# Switching on Superferromagnetism

A. Arora[1, †], L. C. Phillips[2, †], P. Nukala[3], M. Ben Hassine[3], A.A. Ünal[1], B. Dkhil[3], Ll. Balcells[4], O. Iglesias[5] A. Barthélémy[2], F. Kronast[1], M. Bibes[2], and S. Valencia[1,*]

[1]Helmholtz-Zentrum Berlin fur Materialien und Energie, Albert-Einstein-Str. 15, D-12489 Berlin, Germany

[2]Unité Mixte de Physique, CNRS, Thales, Univ. Paris-Sud, Université Paris-Saclay, 91767 Palaiseau, France

[3]Laboratoire SPMS, CentraleSupélec, CNRS-UMR 8580, Université Paris_Saclay, 91190 Gif-sur-Yvette, France

[4]Institut de Ciència de Materials de Barcelona (ICMAB-CSIC), Campus de la U.A.B. 08193 Belltaterra, Spain

[5]Department of Condensed Matter Physics and Institute of Nanoscience and Nanotechnology (IN[2]UB), University of Barcelona, Av. Diagonal 647, 08028 Barcelona, Spain

[†]These authors contributed equally to this work.
[*]Corresponding author: sergio.valencia@helmholtz-berlin.de

**Abstract**

Recent results in electric field control of magnetism have paved the way for the design of new magnetic and spintronic devices with enhanced and novel functionalities and low power consumption. Among the diversity of reported magnetoelectric effects, the possibility of switching on and off long-range ferromagnetic ordering close to room temperature stands out. Its binary character opens up the avenue for its implementation in magnetoelectric data storage devices. Here we show the possibility to locally switch on superferromagnetism in a wedge-shaped polycrystalline FE thin film deposited on top of ferroelectric and ferroelastic $BaTiO_3$ substrate. A superparamagnetic to superferromagnetic transition is observed for confined regions for which a voltage applied to the ferroelectric substrate induces a sizable strain. We argue that electric field-induced changes of magnetic anisotropy lead to an increase of the critical temperature separating the two regimes so that superparamagnetic regions develop collective long-range superferromagnetic behaviour.

The creation of magnetoelectric (ME) systems with coupled electric and magnetic orders[1-3] and in particular the electrical control of magnetism[4-6], is an important research goal with possible applications in spin-based storage and logic architectures with low power consumption[5, 7-9]. Recent work has achieved electrical control of magnetic properties including the type of magnetic ordering[10-12], magnetic moment[13], magnetic anisotropy[14-17], domain structure[18-20], exchange bias[21], spin polarization[22, 23] and critical temperature[10, 24-26]. Where ME effects occur in two-component systems, electrical control of the magnetic phase is achieved either directly by interfacial charge doping, or indirectly via voltage-induced strain transferred by interfacial elastic coupling[5], and the ME behaviour depends crucially on the response of magnetic phases to these charge and strain stimuli.

The interface between ferromagnetic (FM) Fe and ferroelectric and ferroelastic (FE) $BaTiO_3$ (BTO), for instance, garnered interest after charge-induced modulations of magnetic moment[27] and magnetic anisotropy[28] were predicted. While subsequent experiments realized such charge-induced effects at the interface with BTO[12, 14], strain-induced changes of magnetic anisotropy[15, 17] and domain structure[18, 19] appear to be more robust and repeatable.

While most studies of ME effects have focused on thin film-based heterostructured systems[10, 12, 15-17, 19-22, 24-26], electric-field control of nanostructure and nanoparticle properties can also play a role in functionality. To this respect "straintronics", i.e. strain-mediated control of nanoscale magnetic bits in magnetic random access memory devices is emerging as a scalable, efficient, fast and energy efficient alternative to nowadays magnetic memories[29-35]. A recent demonstration of electric-field-control of the blocking temperature of superparamagnetic Ni nanoparticles (NPs) highlighted the possibility to control magnetism at nanometric scales by voltage-induced modification of their magnetoelastic anisotropy[26]. Such nanometric control of magnetism suggests the possibility, not only to modify the magnetic state of a single NP, but also to influence the collective magnetic behaviour of an assembly of NPs. This possibility represents an unexplored extra dimension in ME studies, with potential to unlock novel non-volatile ME behaviours.

Here we explore such ME effects by means of a space resolved analysis of the magnetic domains structure of an ultrathin nanocrystalline Fe film grown on top of a ferroelectric and ferroelastic BaTiO$_3$ substrate. We present evidence that a local superferromagnetic state at room temperature (RT), on otherwise superparamagnetic grains, can result from electrically-induced modification of magnetoelastic anisotropy in a magnetic system coupled to a FE substrate.

## I. RESULTS

### A. Sample description

We present data on an in-situ evaporated Fe film deposited on a 5 mm x 5 mm and 0.5 mm thick BTO substrate. Pseudo-cubic (pc) notation is used to describe crystallographic directions (see Figure 1). A ferroelectric BTO crystal at (and close to) room temperature (RT) has a tetragonal lattice structure (a = 4.036 Å and b = c = 3.992 Å) generally polydomain in nature. The crystal develops a characteristic and well-described pattern with FE polarization oriented always along the elongated c axis. So-called "$a_1$-$a_2$" ferroelastic regions, with domains having polarizations and tetragonal lattice elongation alternating between two orthogonal in-plane substrate directions[11] and domain walls oriented along [110]$_{pc}$, coexist with "$a_1$-$c$" and/or "$a_2$-$c$" regions, with alternating in-plane/out-of-plane polarization, and domain walls oriented along [100]$_{pc}$ and [010]$_{pc}$ directions[36], respectively. While $a_1$ and $a_2$ domains introduce 1.1% uniaxial lattice strain ($a/c$) within the substrate plane c domains are strain isotropic ($b = c$). In the following we introduce notations so to identify FE BTO domains after growth time ($c^{gr}$, $a_1^{gr}$, $a_2^{gr}$), after thermal cycling ($c^T$, $a_1^T$, $a_2^T$) and after voltage application ($c^V$, $a_1^V$, $a_2^V$).

The Fe film is grown (see Appendix) with a ~30-μm wide thickness gradient region (wedge) bisecting the sample, over which the Fe thickness ($t_{Fe}$) varies continuously along the [$\bar{1}$00]$_{pc}$ BTO direction, from 0.5 nm to 3 nm. Elsewhere, the Fe thickness is constant, either "thin" (0.5 nm) or "thick" (3 nm). Space-resolved chemical and magnetic characterization of the sample was performed by means of x-ray photoemission electron microscopy (XPEEM) with tunable energy for x-ray absorption spectroscopy (XAS) and tunable polarization for x-ray magnetic circular dichroism (XMCD) as a magnetic contrast mechanism (see Appendix).

## B. Space-resolved structural, chemical and magnetic characterization

The quality of the wedge was assessed by means of XAS and XMCD right after its growth. Thereafter the sample was *in-situ* capped with a 3 nm Al protective layer and remounted in order to ensure proper electrical contacts for the application of voltage across the sample. The atomic structure of the sample was investigated using a scanning transmission electron microscope (STEM) after the XPEEM measurements.

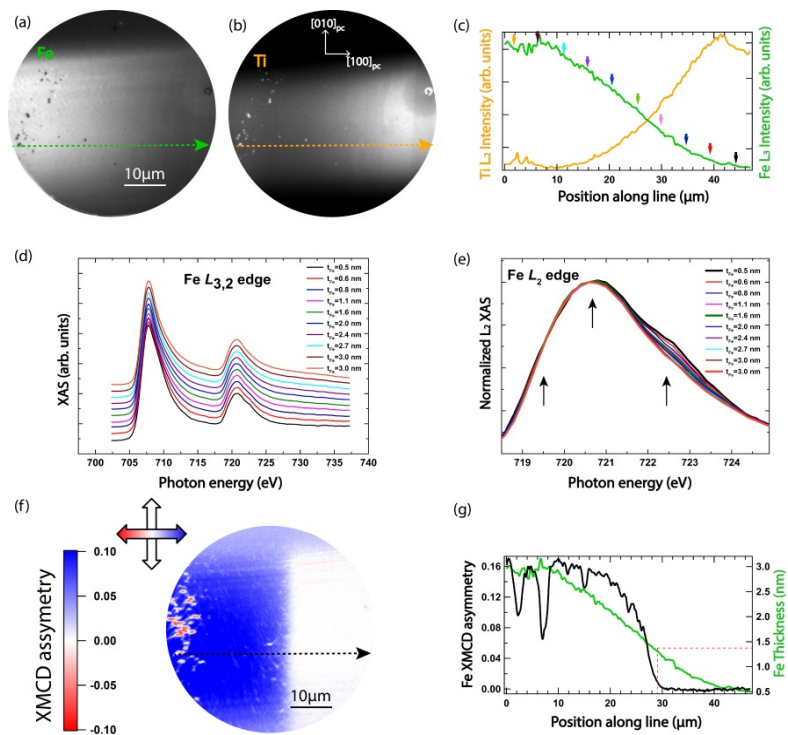

**FIG. 1.** | **Space resolved spectroscopic characterization of the as-grown Fe wedge.** XPEEM images of the Fe wedge obtained at the Fe (a) and Ti (b) $L_3$-edges, respectively, both are normalized to Pre-edge images. c) Fe and Ti intensity profiles measured along dotted lines depicted in panels (a) and (b). d) Fe $L_{2,3}$-edge absorption spectra obtained at different wedge for regions (arrows on panel c) where the Fe thickness varies. Spectra have been normalized at 720.59 eV corresponding to the $L_2$ peak maximum for the thickest Fe region (3.0 nm). Spectra have been shifted vertically for the sake of comparison e) Detail of the Fe $L_2$-edge absorption region showing development of FeO$_x$ characteristic spectral features as $t_{Fe}$ decreases. Spectra are normalized at 720.59 eV. f) Magnetic contrast XMCD image obtained at the Fe $L_3$ edge for the same region depicted in panels (a) and (b). g) Comparison of the profile of the XMCD Vs that of the Fe $L_3$ XAS showing the onset of long-range FM behaviour for $t_{Fe}$ = 1.3 nm. All data taken at 320 K

Normalized elemental XAS images recorded at the Fe $L_3$ (Figure 1(a)) and Ti $L_2$-edges (Figure 1(b)) show opposite gradients of intensity across the wedge, confirming a thickness gradient of Fe (Figure 1(c)) that attenuates the Ti signal from BTO. Spatially-resolved XAS spectra across the Ti $L_{2,3}$-edge (not shown) and Fe $L_{2,3}$- edge (Figure 1(d)) were acquired by scanning the energy of horizontally-polarized incident x-rays. No changes in the Ti $L_{2,3}$ XAS spectral shape is observed across the Fe wedge region within the experimental resolution. On the contrary, the Fe $L_{2,3}$ XAS curves do change as the Fe thickness varies. The evolution is best resolved at the $L_2$-edge spectral region (Figure 1(e)) where changes in the oxidation state of Fe have larger impact in the spectral shape[37]. Thickest Fe regions ($t_{Fe} \approx 3$ nm) show a spectrum akin to that of bulk metallic Fe[37]. As the thickness of Fe is reduced to $t_{Fe} \approx 0.5$ nm, characteristic spectral features of $FeO_x$ (black arrows in Figure 1(e)) emerge and develop although never resembling a pure FeOx spectrum. As the XAS sensitivity to the Fe/BTO interface increases when reducing $t_{Fe}$, this observation highlights the presence of an $FeO_x$ layer at the interface between the film and the BTO substrate. Considering the thickness of the thinnest Fe region and the escaping depth of the electrons on Fe (1.7 Å) the thickness of such FeOx layer should be below 2-3 Å in agreement with the precise FeOx thickness determination by means of STEM[12, 38].

The magnetic domain configuration of the as-deposited Fe wedge is depicted in Figure 1(f). The magnetic contrast XMCD image displays no visible imprint from the underlying FE BTO domain structure expected to arise from the inverse magnetostriction effect[39]. This lack of imprint, expected for films grown at low deposition rates[40] does not exclude certain, although small (below 10% of lattice mismatch), strain transfer between substrate and film[19, 41]. The transition between apparent paramagnetic (white) and ferromagnetic (colored) behavior is sharp as the Fe thickness increases. Comparing a line profile of XMCD across the wedge (Figure 1(g)) with the thickness profile of Fe along the same line reveals a critical Fe thickness $t_{FM} \sim 13$ Å below which long range ferromagnetism is absent. While the existence of $t_{FM}$ is expected from the known suppression of the $T_C$ (and eventual destruction of ferromagnetism) in thin FM films[42], the value of $t_{FM}$ in highly-ordered coherent epitaxial films at 320 K is approximately one atomic monolayer[43], i.e. much lower the present film. Such discrepancy suggest a superparamagnetic behaviour for $t_{Fe} < t_{FM}$ associated with the nanocrystalline structure of the film[44], as discussed below, and/or due to the formation of isolated islands at low coverage [45, 46].

A STEM high-angle annular dark field (HAADF) image (Figure 2), obtained on a section within the region of maximum Fe thickness ($t_{Fe}$ = 3 nm), shows a contiguous and relatively smooth Fe layer. The metal layer shows a nanocrystalline nature with grains of 2-3 nm in size, with the measured interplanar distance on one of these grains as 2.86 Å, corresponding to the lattice parameter of body centered cubic (bcc) Fe.

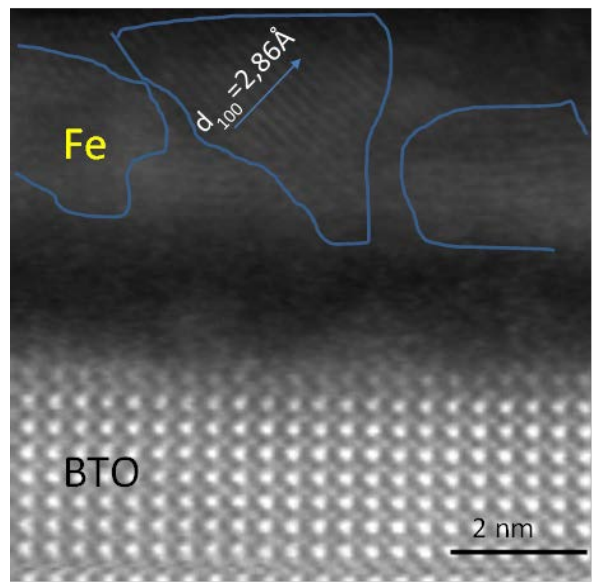

**FIG. 2. | Polycrystalline nature of the Fe film as observed by TEM.** HAADF-STEM image of the heterostructure showing a contiguous nanocrystalline Fe layer. The interplanar distance measured in one of the grains is 2.86 Å, corresponding to the lattice parameter of b.c.c. Fe.

### C. Voltage dependent study of ME effects.

Prior to the application of any electric voltage the sample was cooled down from room temperature to ca. 60 K, and then returned to 320 K. This temperature cycle altered the initial magnetic domain structure of Fe as the stress changes suffered by the BTO substrate during this thermal excursion are expected to be fully transferred to the top metallic layer[47] via inverse magnetostriction effect[39].

Figure 3(a) depicts the XMCD image obtained at V = 0 V on a representative region (different from that of Figure 1). A striped magnetic domain pattern with magnetic domain walls oriented along $[\bar{1}10]_{pc}$, is visible within the FM (thick) side of the wedge. The magnetization direction, within these domains, points alternately along $[010]_{pc}$ / $[0\bar{1}0]_{pc}$ (new white stripes) and $[100]_{pc}$ directions (original blue regions), see Supplemental Material (S.M.) section 1 for details. This FM domain pattern, with new FM regions where the magnetization direction has rotated by 90° from its initial orientation could be, depending on the FE domain pattern at growth time, an indirect manifestation of an underlying "$a^T_1$-$a^T_2$" (refs. 19, 36, 41, 48) or purely "$c^T$" (refs. 19, 48) ferroelastic texture after the thermal cycle (T).

The application of an out-of-plane voltage $V$ = +74 V across the substrate thickness leads initially to a better definition (see Figure 3(b)) of the FM domain-stripped pattern shown in Figure 3(a). After waiting for ~1 h (holding the voltage) new FM domains, with magnetization direction either along $[100]_{pc}$ (blue XMCD) or alternating $[010]_{pc}$ / $[0\bar{1}0]_{pc}$ (white XMCD contrast) become visible, see Figure. 3(c) and S.M. Figure. S1. The new voltage (V)-induced FM domain walls extend along $[100]_{pc}$ resembling ferroelastic $a_1$-$c$ walls, hence we conclude that the underlying BTO has transformed to $a^V_1$-$c^V$ FE pattern. This result excludes a "$c^T$" FE domain within the observed area after the thermal excursion[19, 48] and points to an $a^T_1$-$a^T_2$ ➔ $a^V_1$-$c^V$ FE domain transformation[11, 20, 36, 40] (see S.M.).

We label the FM regions observed at V = +74 V according to the orientation of their magnetization and domain walls. Stripe-shaped magnetic regions with domain walls along the $[100]_{pc}$ axis and homogeneous XMCD contrast in Figure 3(c) have been labelled $α$ and $β$ depending on whether the magnetization direction is aligned along $[010]_{pc}$ / $[0\bar{1}0]_{pc}$ (white-colored XMCD) or along $[100]_{pc}$ (blue-colored XMCD), respectively. Regions for which the magnetic imprint of former ferroelastic $a^T_1$-$a^T_2$ domains persists has been labelled as $γ$. Further increasing the voltage (up to +170 V) leads to the movement and occasional interchange of FM domain regions, see Figure. 3(d). Although the coexistence of $α$-, $β$- and $γ$-like FM regions has not been reported yet, its formation can be explained based on a differential strain model[40, 48] which requires an $a^{gr}_1$-$c^{gr}$ FE domain pattern underneath the metallic film at growth (gr) time which transforms into $a^T_1$-$a^T_2$ after the thermal cycle (see S.M. Section 3).

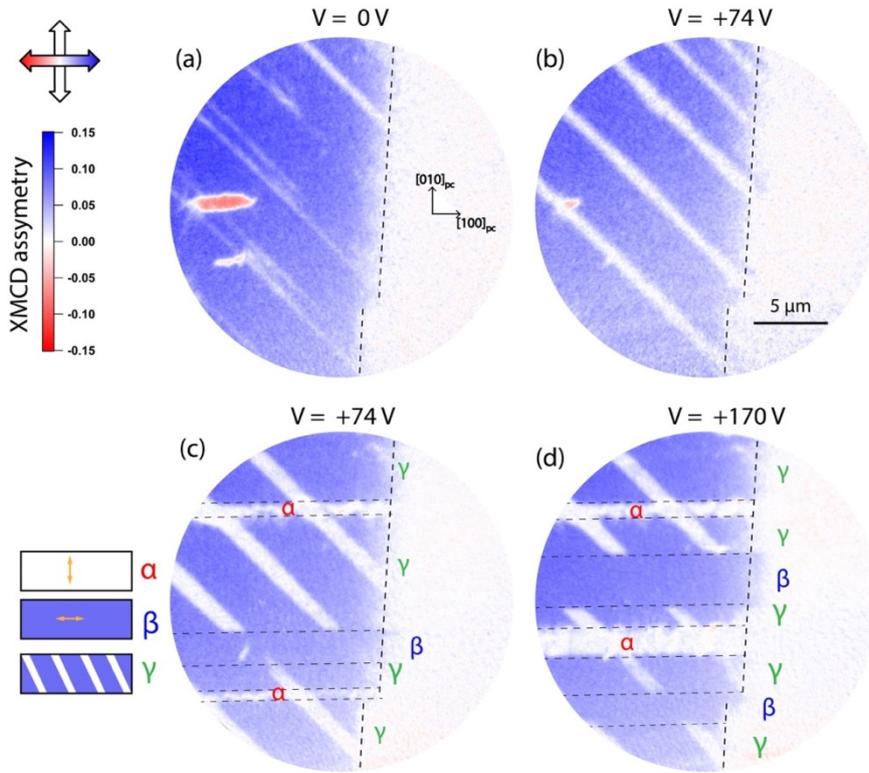

**FIG. 3. | Voltage dependent evolution of magnetic domains on Fe.** XMCD PEEM images obtained at the Fe $L_3$-edge at various voltages (a) 0 V, (b) and (c) 74 V and (d) 170 V, respectively. Images depicted at panels (b) and (c) where obtained with a time difference of ca. 2h. All images were obtained at RT after a thermal excursion to low temperature. Image region is not the same as that depicted in Figure 1. α-, β- and γ- FM domains are delimited by a dashed line along $[010]_{pc}$

Despite the thickness gradient of the Fe layer, no noticeable dependence of the new magnetic domain pattern is seen along $[100]_{pc}$ axis. However, we do observe that the application of a voltage has led to the extension of long-range FM behaviour towards lower Fe thicknesses at localized β regions only (see Figure. 3(c, d). As the incoming photon beam is orthogonal to the $[010]_{pc}$ direction the XMCD lacks magnetic sensitivity for α domains. An azimuthal rotation of the sample by 45° leads to magnetic sensitivity to both $[100]_{pc}$ and $[010]_{pc}$ magnetic axis, and consequently to both α and β FM regions. Further voltage cycling (from +170 V to -170 V) demonstrates that this extension of long-range FM order to lower $t_{Fe}$ appears indeed only at β regions. Some of these extended regions contract back after some time, as seen for example by comparing Figure. 3(d) and Figure. 4(a). This is indeed expected for polycrystalline materials as a time dependent relaxation of strain may take place after modification of the ferroelectric and ferroelastic domain configuration[49]. Only those Fe regions where the local BTO domain has transformed (by FE domain wall motion) more recently retain the extended long-range FM ordering. This can be seen by comparing panels (b) and (c) of Figure 4 where dash lines along $[100]_{pc}$ delimit regions which have recently undergone an α to β transition. The analysis of the extended FM regions depicted in Figure. 4 (d) shows that the growth of long-range magnetic order extends ca. 1.3 µm along $[100]_{pc}$ which corresponds to a decrease of ∼ 1 Å for the critical thickness for FM for our polycrystalline Fe film, see S.M. section 5 for details.

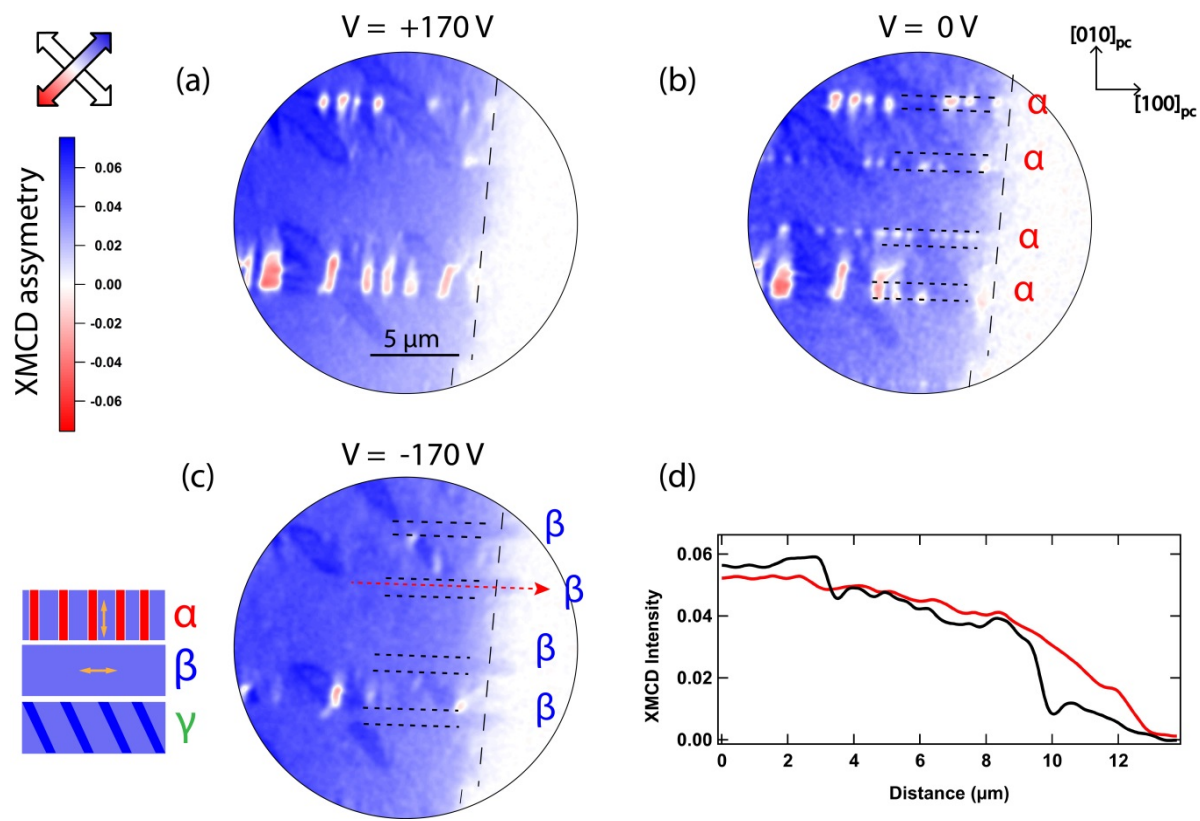

**FIG. 4. | Extension of long range FM ordering towards lower $t_{Fe}$ regions upon application of V.** (a), (b) and (c) XMCD PEEM images obtained at the Fe $L_3$-edge after a 45° azimuthal rotation of the sample for decreasing voltages, i.e. (a) V = 170 V, (b) V = 0 V and (c) V = -170 V. The azimuthal rotation of the sample allows XMCD to be sensitive to magnetization oriented along both $[100]_{pc}$ and $[010]_{pc}$. (d) Line profile of the XMCD along red line shown in panel (c) for V = -170 V (black) and V = 170 V (red). Long range ferromagnetism has extend ca. 1.3 µm along $[0\bar{1}0]_{pc}$ after a localized α to β magnetoelastically triggered transition. Such extension corresponds to an decrease of the onset of FM ($t_{FM}$) of ca. 1 Å.

## II. DISCUSSION

The lack of direct information about the ferroelastic domain configuration of BTO complicates the interpretation of the experimental results. However, the analysis of the XMCD vs voltage data within a differential strain model[40, 48] makes it possible to work out the sign of the effective magnetostriction for our system (see S.M. Section 2), understand the localized rotation of magnetization orientation as function of V in terms of changes of the magnetoelastic anisotropy, and identify the origin of α, β and γ FM domains in terms of magnetoelastic coupling history (see S.M. Sections 3 and 4). To this respect, it can be shown that for large V (FE "$c^V$" domain configuration), and relative to the strain state at growth time, α regions and γ regions are unstrained, while β regions are under an uniaxial lattice strain compression of ca. 1% along $[010]_{pc}$. This strain state, which arises from FE $a_1^{gr}$ domains at growth time which transforms into a $c^V$ ones when voltage is applied, leads to a magnetic easy axis along $[100]_{pc}$.

Recent results have shown the possibility to control the antiferromagnetic/FM metamagnetic transion temperature of FeRh deposited on top of BTO by electric-field modification of its unit cell volume[10]. Doing an analogy, our observed extension of the onset long-range FM behaviour (towards regions where magnetic ordering was previously absent) could be interpreted as a local increase of the ferromagnetic transition temperature ($T_C$) due to voltage-induced changes in the unit cell volume of Fe. To address this possibility we have checked the required change in temperature necessary to observe an extension of long-range FM ordering as that depicted in Figs. 4 (c) and (d) and compared it with the expected strain-dependency of $T_C$ for Fe. Our analysis, reported on S.M. section 6, leads to the result that an increase of $T_C$ of about 50 K would be necessary to lead to the observed extension of long-range FM ordering. Such a change of $T_C$ cannot originate from a voltage-induced Fe bcc unit cell volume modification of Fe. Experimental and theoretical results have evidenced an almost lack of sensitivity of $T_C$ of Fe bcc towards elastic strain for values such as those provided by the BTO substrate[50-52]. At this point we recall the polycrystalline nature of our film as well as the surprising absence of long-range ferromagnetic ordering for $t_{Fe} < 13$ Å. We note that for Fe/BTO it has recently been reported that superparamagnetism does arise due to the formation of non-connected Fe nanoislands in the thickness range of 1- 4 monolayers[12, 14] due to a Volmer-Weber growth[45, 53]. We hence consider that at room temperature and for $t_{Fe} < 13$ Å our film consists of isolated nanoscale grains behaving as nanoparticles with a superparamagnetic limit which depends on grain size and magnetic anisotropy, the latter susceptible to be modified by an electric field (or a voltage). Indeed electric field enhancement of the Blocking temperature ($T_B$) of non-interacting NP by ~40 K has recently been reported[26]. Within this work Kim *et al* show that compressive strain increases the magnetoelastic contribution to the magnetic anisotropy and enhances the magnitude of the energy barrier for spin flip stabilizing the magnetization direction of the macrospin associated to the NP at RT[26]. However this magnetoelectric effect does not explain our results. Indeed, within an assembly of NP the magnetization direction of the individual NPs will point along one of the energetically equivalent directions defined by the magnetic anisotropy axis. In such a case, and due to the limited spatial resolution of XPEEM (~30 nm) as compared to the NP grain size (~3 nm), the XMCD would average to zero[14]. On the contrary, when we apply the voltage, regions under compressive strain do show a sizable XMCD signal thus, pointing to a long-range ferromagnetic-like behaviour. This collective magnetic ordering is characteristic of superparamagnetic NP with strong interparticle interactions which undergo a magnetic transition to a superferromagnetic state at certain temperature[54-61] $T_P$. A SPM to a super spin glass transition can be excluded as the frustration and the randomness characteristic of this magnetic phase[55, 59] would lead to zero averaged magnetization and hence to XMCD = 0, in contradiction with the data depicted in Figure 4 and Figure S8. The transition temperature between the superparamagnetic (T > $T_p$) and superferromagnetic state (T < $T_p$) depends on both, interparticle interactions and magnetic anisotropy. Hence it is expected that, as for the case of non-interacting NP[26], voltage controlled changes of the magnetic anisotropy via magnetoelastic coupling in NP systems with strong interparticle interactions can also lead to a shift of its associated superparamagenetic/superferromagnetic (SPM/SFM) transition temperature $T_p$[54, 59]. We thus conclude that β regions showing an extension of long-range FM ordering have undergone a SPM/SFM transition as the applied voltage has modified the strain conditions and the associated magnetic anisotropy so to increase $T_p$ by ca. 50 K (see S.M.) above 320 K.

To test the plausibility of this hypothesis we have modelled the nanocristalline nature of the sample by an assembly of 20 X 20 NPs and computed their average magnetization as function of temperature and interparticle distance (r) by means of Monte Carlo simulations. We have considered two cases i) strain free NPs and ii) NPs subject to a uniaxial lattice strain compression of 1%. For the sake of keeping computation time within reasonable limits we have modelled the NPs as spheres of diameter D=2.5 nm with macrospins $S_i$ occupying the nodes of a regular triangular lattice and having a magnetic moment proportional to their volume (V) and saturation magnetization ($M_s$). The interaction of Heisenberg spins has been accounted by considering the following Hamiltonian

$$H = -K_c V (\vec{S}_i \cdot \hat{n}_i)^2 + g \sum_{i<j} \left( \frac{\vec{S}_i \cdot \vec{S}_j - 3(\vec{S}_i \cdot \hat{r}_{ij})(\vec{S}_j \cdot \hat{r}_{ij})}{|\vec{r}_{ij}|^3} \right) \quad (1)$$

where the first term of Eq. 1 stands for uniaxial anisotropy, being $\hat{n}_i$ a local random anisotropy direction, and the second one accounts for interpaticle dipolar interactions. All quantities in parenthesis are dimensionless. When distances between NPs are measured in terms of particle diameter as $r = \rho \cdot D$, the dipolar energy strength is controlled by the parameter $g$ given by

$$g = \frac{\mu_0 \pi M_s^2 D^3}{24 \rho^3} \quad (2)$$

For the case of close contact ($\rho = 1$) this parameter amounts to 725 K while the anisotropy energy is of the order of 225 K for the usual values of Fe $K_c$= 4.8 10$^4$ J/m$^3$ and $M_s$= 1.71 10$^6$ A/m (ref. 62)

Simulations have been run starting from high temperature with disordered magnetic moments and decreasing T in constant steps in absence of magnetic field down to 1 K while averaging the magnetization components. Results of the thermal dependence of the magnetization are shown in Figure 5 for $\rho = 1.5$, 1.75 and 2. In spite of the noise at high temperature due to finite size effects our simulations show that the magnetization of the ensemble of NPs becomes non-zero below a temperature ($T_P$) that depends on the interparticle distance, and increases as T further decreases. This collective ferromagnetic-like ordering is driven by dipolar interactions between the NPs.

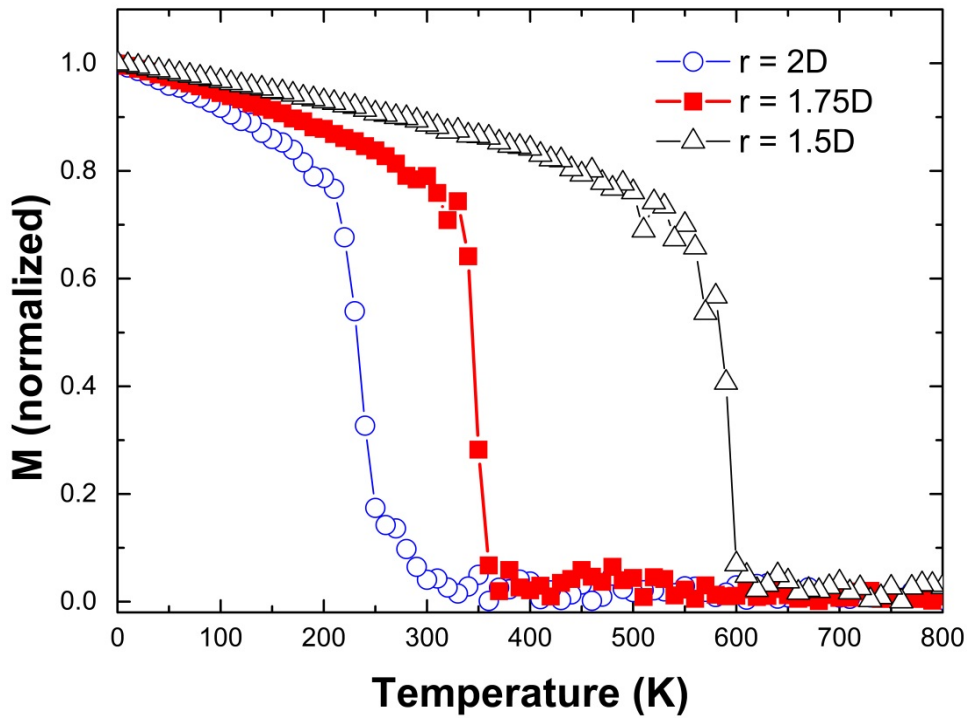

**FIG. 5. | Results of Monte Carlo simulations of an ensemble of dipolar interaction NPs.** Thermal dependence of the magnetization for interparticle distances $\rho=$ 1.5, 1.75, 2 (in multiples of the particle diameter) from right to left.

The effect of a voltage induced compression of $\beta$ regions by ca. $\bar{\varepsilon}^{FM} = -1\%$ has been accounted for by including an effective increase of the NPs anisotropy due to magnetoelastic coupling by an amount $K_{me}= 3/2\ Y\lambda\bar{\varepsilon}^{FM}$, being $Y$ de Young modulus and λ the magnetostriction coefficient. Values for Y and λ for Fe have been obtained from ref. 39. Figure 6 shows a comparison of the temperature dependent magnetization for an ensemble of NPs with $\rho$=2 having i) random uniaxial magnetocrystalline anisotropy (black circles) and ii) an additional uniaxial term due to the voltage induced magnetoelastic anisotropy $K_{me}$ (red squares). The results of the simulation show that the electric-field induced strain induces a sizable FM-like long range order of NPs macrospins in a range of temperatures for which the unstressed ensemble was SPM (signalled by the rectangular dashed area).

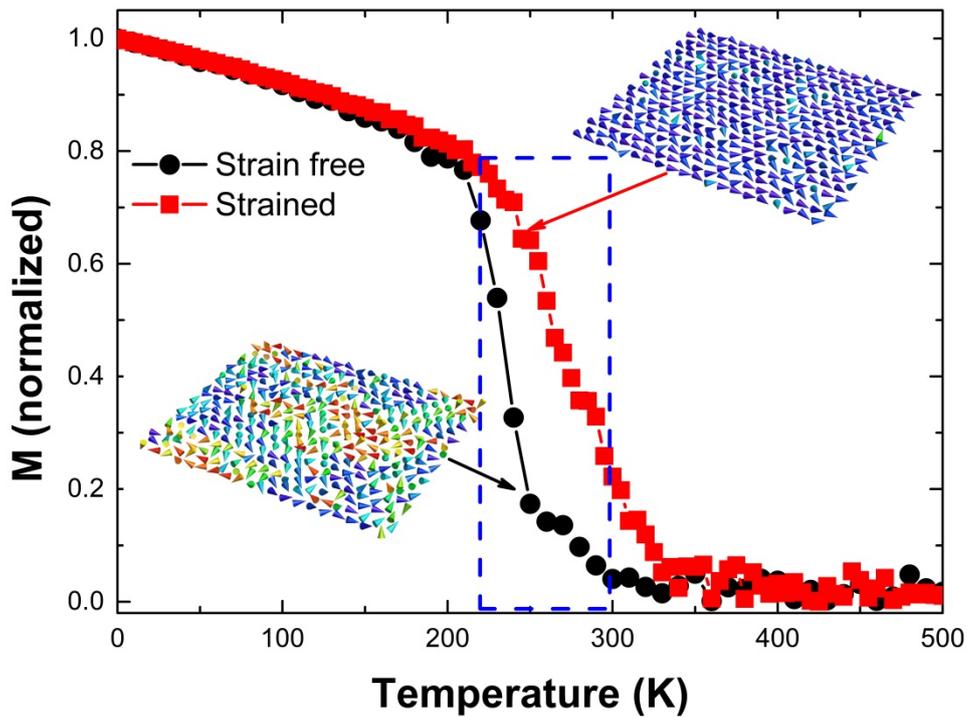

**FIG. 6 | Effect of electric-field induced stress on transition temperature $T_P$ between SPM and SFM state.** Thermal dependence of the magnetization as obtained by simulations of an ensemble of dipolar interacting NP on a triangular lattice with interparticle distance $\rho$= 2. Black-circles curve correspond to an ensemble of NPs with random magnetocrystalline anisotropy. Blue-triangles curve takes into account an additional magnetoelastic contribution to the anisotropy ($K_{me}$) resulting from an electric-field induced compression of 1%. The dashed rectangle marks the region where long range FM order is induced by the strain induced anisotropy. Left (right) inset displays representative spin configuration corresponding to circle (square) curves for a temperature inside the rectangular region.

The change in the $T_P$ as estimated from the inflection points of both curves is of the order of $\Delta T_c \approx (40\pm10)$ K in agreement with our experiments.

### III. CONCLUSION

Application of electric-field induced strain to a polycrystalline Fe wedge sample shows localized extension of long range ferromagnetism towards lower Fe thicknesses. This result can be explained by a magnetoelastic modification of the magnetic anisotropy associated to the nanometer sized Fe crystals inducing a local superparamagnetic/superferromagnetic transition at room temperature. Our results open up new possibilities for electric field control of magnetic ordering as the size of the superferromagnetic domains can be controlled and reduced by scaling down the size of FE domains by using thin films instead of single crystals[63].

## Acknowledgements

We thank HZB for the allocation of synchrotron radiation beamtime. P.N., M.B.H., A.B., M.B. and B.D. thank a public grant overseen by the French National Research Agency (ANR) as part of the "Investissements d'Avenir" program (Reference:ANR-10-LABX-0035, LabexNanoSaclay). P.N., M.B.H. and B.D. aknowledge the MATMECA project supported by the ANR under contract number ANR-10-EQPX-37. Ll.B. thanks MEIC through the Severo Ochoa Programme for Centres of Excellence in R&D (SEV-2015-0496) and MAT2015-71664-R, FEDER, EU Horizon 2020 under the Marie Sklodowska-Curie grant agreement No.645658. O.I. acknowledges financial support from the Spanish MINECO (MAT2015-68772), Catalan DURSI (2014SGR220) and European Union FEDER Funds, also CSUC for supercomputer facilities.

## APPENDIX: METHODS

***Sample preparation.*** The films were deposited onto the (001) surface of a $BaTiO_3$ crystal (5 mm × 5 mm × 0.5 mm, *SurfaceNet GmbH*) by electron-beam-assisted evaporation, in vacuum and at room temperature. The substrate contained a mixed ferroelectric domain structure, and was not annealed before deposition. A knife edge positioned close to the substrate shielded approximately half of the substrate area from the Fe deposition. The deposition rate was 0.1 nm/min in the thick region away from the wedge, and proportionally less elsewhere. After wedge deposition (2.5 nm at the thicker region) the blade was withdrawn, and a further 0.5 nm of Fe and 3 nm of Al were deposited uniformly over the substrate. The resulting wedge extends along a [100] direction of BTO, where the Fe thickness varied from (3 ± 0.3) nm to (0.5 ± 0.1) nm over a lateral distance of approximately 50 µm. After deposition, the sample was transferred in situ to the XPEEM chamber without breaking vacuum for chemical and magnetic characterization. Afterwards 3nm of Al were deposited in-situ so to avoid oxidation when removing the sample in order to fix electrical contacts for the XPEEM experiments as function of V.

**STEM and TEM.** Electron transparent sample cross-sections were prepared via focussed ion-beam based sample preparation procedure, followed by low kV (2 V and 1 V) cleaning steps. High angle annular dark-field STEM was performed several months after PEEM measurements at 200 kV on a Titan G2, $C_s$ corrected microscope.

**XPEEM, XAS and XMCD measurements.** High-resolution images were taken at the spin-resolved photo-emission electron microscope at the synchrotron radiation source BESSY II operated by the Helmholtz-Zentrum-Berlin. This set-up is based on an Elmitec III instrument with an energy filter, permanently attached to an undulator beamline with full polarization control, in an energy range from 80 to 2,000 eV. The lateral resolution of the spin-resolved photo-emission electron microscope is about 25 nm for X-ray excitation. Space resolved spectroscopic information was obtained by PEEM acquiring images at different photon energies of an incoming (horizontal) linearly polarized beam across the Fe $L_{2,3}$ edges with 0.2 eV steps. After normalization to a bright-field image, the sequence was drift-corrected and spectra were obtained at specific thickness of the Fe wedge as indicated in Figure 1 (d) and (e). Normalized elemental XAS images depicted in Figure 1(a) and (b) were acquired by dividing the PEEM signal on the peak of the Fe-edge (707.8 eV) or the Ti-edge (457.8 eV) by a pre-edge image taken at a slightly lower X-ray energy, 700 eV and 450 eV in case of Fe and Ti, respectively. For magnetic imaging the photon energy was tuned to the $L_3$ resonance of metallic iron (707.8 eV) to exploit the element-specific XMCD. Each of the XMCD images shown was calculated from a sequence of 80 images (3s each) taken with circular polarization (90% of circular photon polarization) and alternating helicity. After normalization to a bright-field image, the sequence was drift-corrected, and frames recorded at the same photon energy and polarization has been averaged. The Fe magnetic contrast is shown as the difference of the two average images with opposite helicity, divided by their sum. The magnetic contrast represents the magnetization component pointing along the incidence direction of the X-ray beam, and we study different in-plane components by rotating the sample azimuth in the beam. Voltages *V* were applied to a bottom contact created using conducting silver paint, where the film surface was held at constant potential, to create electric field across the thickness of the BTO substrate.

**Author contributions**

S.V. and M.B. designed and conceived the experiment. L.P., S.V., A.A.Ü, F.K. and S.V. were involved in the PEEM experiments. M.B.H., P.N. and B.D. performed electron microscopy. O.I. performed the Monte Carlo simulations. A.A., L.P. and S.V. analyzed and interpret the data with inputs from M.B., Ll.B. and A.B. A.A., L.P. and S.V. wrote the manuscript with inputs from all co-authors.